\documentclass{ws-ijmpd}

\begin{document}
  
\markboth{Zden\v{e}k Stuchl\'{i}k and Ji\v{r}\'{i} Kov\'{a}\v{r}}
{Pseudo-Newtonian gravitational potential for Schwarzschild-de Sitter spacetimes}

\catchline{}{}{}{}{}
%
%%%%%%%%%%%%%%%%%%%%%%%%%%%%%%%%%%%%%%%%%%%%%%%%%%%%%%%%%%%%%%%%%%%%

\title{PSEUDO-NEWTONIAN GRAVITATIONAL POTENTIAL\\FOR SCHWARZSCHILD-DE~SITTER SPACETIMES}

\author{ZDEN\v{E}K STUCHL\'{I}K and JI\v{R}\'{I} KOV\'{A}\v{R}\dag}

\address{Institute of Physics, Faculty of Philosophy and Science, 
Silesian University in Opava\\Bezru\v{c}ovo n\'{a}m. 13, Opava, CZ-74601,
Czech Republic\\
\dag Jiri.Kovar@fpf.slu.cz}

\maketitle

\begin{history}
\received{Day Month Year}
\revised{Day Month Year}
\end{history}

\begin{abstract}
Pseudo-Newtonian gravitational potential describing the gravitational field of static and spherically symmetric black holes in the universe with a repulsive cosmological constant is introduced.
In order to demonstrate the accuracy of the pseudo-Newtonian approach, the related effective potential for test-particle motion is constructed and compared with its general relativistic counterpart given by the Schwarzschild-de~Sitter geometry.
The results indicate that such an approach could be useful in applications of developed Newtonian theories of accretion discs in astrophysically interesting situations in large galactic structures for the Schwarzschild-de~Sitter spacetimes with the cosmological parameter $y=(1/3)\Lambda M^2\leq10^{-6}$.
\end{abstract}


%************************************************************************
\section{Introduction}
\label{section 1}
Recent cosmological observations of distant supernova Ia type explosions suggest an accelerating universe; starting at the cosmological redshift $z\sim1$\cite{Spe-etal:2003,Spe-etal:2007} the accelerated expansion should be generated by the so-called dark energy. These results are in accord with detailed observations of the anisotropies of the microwave cosmic background radiation indicating that the dark energy represents $\sim 73\%$ of the energy content in the observable universe\cite{Bah-etal:1999}; the sum of energy densities is very close to the critical energy density $\rho_{\rm crit}$ corresponding the parabolic universe in concordance with the inflationary cosmological model\cite{Bah-etal:1999,Wan-eta:2000}.
There is a variety of possible candidates for the dark energy\cite{Cop-Sam-Tsu:2006,Alc:2006}. Most of the related models have the equation of state with parameter $w=p/\rho$ varying during the cosmological expansion. 

Nevertheless, the recent observational data indicate that the allowed equation of state is very close to the case of constant $w=-1$ corresponding to the repulsive cosmological constant $\Lambda>0$ with $\rho_{\Lambda}\sim0.73\rho_{\rm crit}$\cite{Spe-etal:2003}.  Therefore, it is quite important to consider the influence of $\Lambda>0$ in astrophysically relevant situations quite seriously. 

We have tested possible effects of $\Lambda>0$ in astrophysically motivated problems\cite{Stu:2005}, investigating namely the properties of black-hole solutions of Einstein equations for the test-particle motion in Schwarzschild-de Sitter (SdS)\cite{Stu-Hle:1999} and Kerr-de Sitter (KdS)\cite{Stu-Sla:2004} spacetimes and the test perfect fluid in SdS\cite{Stu-Sla-Hle:2000} and KdS\cite{Sla-Stu:2005} spacetimes. Further, the test-particle and fluid properties were treated also in the framework of the optical reference geometry\cite{Kov-Stu:2006,Kov-Stu:2007}  allowing introduction of inertial forces in the intuitively natural Newtonian way\cite{Abr-Car-Las:1988,Abr-Nur-Wex:1995}.

The influence of the repulsive cosmological constant on the black-hole spacetime structure can be properly represented by the dimensionless cosmological parameter $y=(\Lambda/3)M^2$. For SdS black holes admitting existence of stable circular geodesics, i.e., existence of accretion discs, the cosmological parameter $y<y_{\rm ms,e}\doteq0.000237$ \cite{Stu-Hle:1999}. The cosmological tests using the supernova magnitude-redshift relation and the Cosmic Microwave Background Radiation fluctuations measurements\cite{Spe-etal:2003,Spe-etal:2007} imply $\Lambda\approx10^{-56}$cm$^{-2}$, and thus very low values of $y$ for astrophysically realistic black holes. In fact, $y\sim10^{-24}$ for super-massive black holes\cite{Stu-Sla-Hle:2000}; strong optical observable effects could be expected for super-giant black holes\cite{Ish-etal:2007} (or clusters of galaxies) with $M\geq10^{15}M_{\odot}$ \footnote{For primordial black holes in the very early universe, with expected high values of effective cosmological constant, the values of $y$ can be much closer to $y_{\rm ms,e}$. Considering the electroweak phase transition at $T_{\rm ew}\sim100$GeV, we obtain an estimate of the primordial effective cosmological constant $\Lambda_{\rm we}\sim0.028$cm$^{-2}$, while at the level of the quark confinement at $T_{qc}\sim1$GeV we obtain $\Lambda_{\rm qc}\sim2.8\times10^{-10}$cm$^{-2}$ and consequently higher values of $y$ \cite{Stu:2005,Stu-Sla-Hle:2000}.}. 

For astrophysically realistic SdS black holes, the strong gravity near the black hole horizon $r_{\rm h}\sim2M$ weakens with distance growing and at $r>>M$ can be described quite well by the Newtonian theory. However, the Newtonian theory looses its validity near the so-called static radius $r_{\rm s}\sim y^{-1/3}M$, where the repulsive effect of the cosmological constant starts to be relevant up to the other strong gravitation region  near the cosmological horizon $r_{\rm c}\sim y^{-1/2}M$. Therefore, the cosmological constant has relevant influence on the structure of disc configurations (both geometrically thin and thick) introducing quite naturally outer edge of the accretion discs\cite{Stu:2005,Stu-Sla-Hle:2000}.

It should be stressed that all of the relevant effects of $\Lambda>0$ on the black-hole structure are quite well expressed in the SdS spacetimes, since the rotational effects of the black hole spin are concentrated into the region in close vicinity of the black-hole horizon, where the influence of $\Lambda>0$ can be abandoned for realistic values of the black hole mass and the relict cosmological constant\cite{Stu:2005}. Of course, the efficiency of the accretion processes is controlled by the rotational effects in the innermost parts of the disc, where the KdS spacetime structure is relevant\cite{Stu:2005}. For near-extreme black holes the efficiency takes large values $\sim0.4$ as compared with the SdS spacetimes where the efficiency $\sim0.059$. Therefore, in studying the large scale properties of disc structures, investigation of the SdS spacetime is quite sufficient, only the accretion efficiency has to be given by the K(dS) spacetime structure, governing the innermost parts of the disc.

For this reason, it is worth to realize more detailed studies of the disc structures around supermassive black holes reflecting the influence of the repulsive cosmological constant. There is one especially important problem that could be hardly solved in the framework of full general relativistic approach, namely influence of $\Lambda>0$ on the structure of self-graviting disc. We expect that the pseudo-Newtonian approach could be successful. For such purposes, we introduce here a pseudo-Newtonian potential $\psi$ of the SdS spacetimes that could enable us to use directly standard techniques developed in the framework of Newtonian physics, e.g., the Newtonian discoseismology\cite{Kat-Fuk-Min:1998}, taking the influence of the repulsive cosmological constant and some relativistic effects into account. 

There is a variety of different approaches in defining the pseudo-Newtonian gravitational potential describing black holes and various aspects of their spacetime structure\cite{Pac-Wii:1980,Cha-Kha:1992,Now-Wag:1991,Art-Bjo-Nov:1996,Sem-Kar:1999,Muk:2002,Muk-Mis:2003,Gho-Muk:2007}. In the case of Schwarzschild spacetimes, it seems\cite{Art-Bjo-Nov:1996} that to reflect the accretion disc properties, the most convenient is the Paczynski-Wiita (P-W) gravitational potential\cite{Pac-Wii:1980} $\psi_{\rm P-W}=-1/(r-2GM/c^2)$. It enables us to calculate positions of the marginally stable and bound circular orbits at the same radii as follow from the general relativistic calculations. 

Originally, the P-W potential was introduced by a guess, when attempting to include the Schwarzschild radius into the Newtonian gravity. However, there is a simple heuristic method for derivation of the pseudo-Newtonian potentials that yields the P-W potential. The same method was used for the derivation of the pseudo-Newtonian gravitational potential for the equatorial plane of Kerr spacetimes as well\cite{Muk:2002}. Then the position of the marginally stable circular orbit  corresponds to the position determined by using the general relativistic approach, and differences in positions of marginally bound circular orbit determined in both the ways are relatively small. In this paper, this heuristic method is just used for derivation of the pseudo-Newtonian gravitational potential for the SdS spacetimes.

Defining such potential, we have to reflect properly both the gravitational attraction of the black hole and the repulsive effects of the cosmological constant. The P-W potential, describing with high precision properties of accretion processes in the field of Schwarzschild black holes has naturally defined behaviour at infinity. In the SdS spacetimes, we have to find a proper reference point that could in a well defined way serve in a similar sense as infinity in asymptotically flat spacetimes. It is shown\cite{Stu-Hle:1999,Hle:2002} that such a role could be attributed to the so-called static radius, where the gravitational attraction of black hole acting on matter is exactly balanced by the cosmic repulsion, i.e., test-particles feel no force, similarly to asymptotic infinity in flat spacetimes. Concentrating on properties of stationary disc configurations in the SdS spacetimes, the static radius is appropriately chosen; we do not consider here properties of the spacetime near the cosmological horizon of the SdS spacetimes, nevertheless, the pseudo-Newtonian potential we are using here reflects the basic spacetime property near the cosmological horizon, since it diverges there.

In the pseudo-Newtonian potential we properly fix position of the marginally stable orbit, marginally bound orbit, the static radius, and we establish the particle energy at these radii in agreement with the exact general relativistic values. 
%*************************************************************************
\section{SdS spacetime and its pseudo-Newtonian gravitational potential}
\label{section 2}
In the standard Schwarzschild coordinates $(t,r,\theta,\phi)$, and the geometric system of units $(c=G=1)$, the SdS spacetimes are determined by the line element\cite{Stu-Hle:1999}
\begin{eqnarray}
\label{1}
ds^2=&&-\left(1-\frac{2M}{r}-\frac{\Lambda}{3}r^2\right)dt^2\nonumber\\&&+\left(1-\frac{2M}{r}-\frac{\Lambda}{3}r^2\right)^{-1}dr^2+r^2\left(d\theta^2+\sin^2{\theta}d\phi^2\right),
\end{eqnarray}
where $M$ is the mass parameter of the spacetimes. It is useful to introduce the dimensionless parameter $y=\Lambda M^2/3$ and use dimensionless coordinates $t\rightarrow t/M$, $r\rightarrow r/M$, which is equivalent to putting $M=1$.

\begin{figure}[tpb]
\centerline{\psfig{file=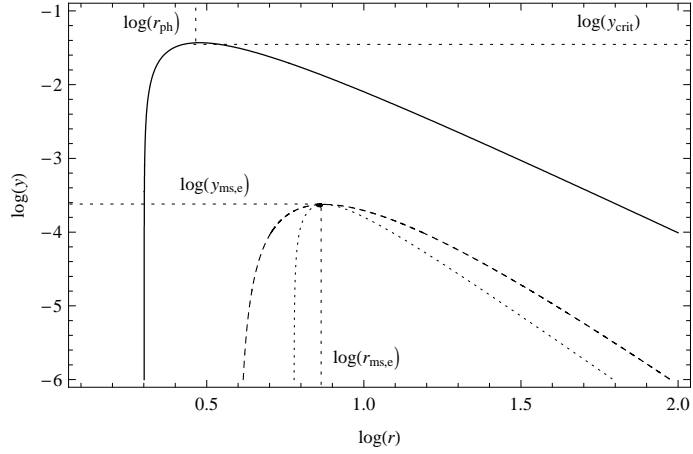,width=0.75\hsize}}
\vspace*{8pt}
\caption{\label{Fig:1}Functions $y_{\rm h}(r)$ (solid) determining radii of horizons, $y_{\rm mb}$ (dashed) determining marginally bound circular orbits, and $y_{\rm ms}(r)$ (dotted) determining marginally stable circular orbits. Regions above the solid curve correspond to dynamic parts of SdS spacetimes.}
\end{figure}

Singularities of the line element, i.e.,  black-hole and cosmological horizons, are given by the relation $1-2/r-yr^2=0$, thus by solutions of the equation 
\begin{eqnarray}
\label{2a}
y=y_{\rm h}\equiv\frac{r-2}{r^3},
\end{eqnarray}
which can be expressed in the form 
\begin{eqnarray}
r_{\rm h}=\frac{2}{\sqrt{3y}}\cos{\frac{\pi+\xi}{3}},\quad
r_{\rm c}=\frac{2}{\sqrt{3y}}\cos{\frac{\pi-\xi}{3}},
\end{eqnarray}
where $\xi=\cos^{-1}{(3\sqrt{3y})}$. 
Both the horizons exist for $0<y<y_{\rm crit}=1/27$, separating the spacetimes into two dynamic and one static regions. For $y=y_{\rm crit}$, both the horizons coalesce at the radius of the photon circular geodesic at $r=r_{\rm ph}=3$. For $y>y_{\rm crit}$, the horizons disappear and the SdS spacetimes become dynamic naked-singularity spacetimes (see Fig.~\ref{Fig:1}). 

The heuristic method\cite{Muk:2002} enabling us to define the P-W gravitational potential and the pseudo-Newtonian gravitational potential for the equatorial plane of the Kerr spacetimes is based on the knowledge of exact general relativistic relations for the angular momentum per particle mass $L_{\rm c}$ and energy per particle mass $E_{\rm c}$ of the circular geodesics. The Newtonian gravitational potential $\psi_{\rm N}$ for central gravitational fields is related to the Newtonian angular momentum per particle mass $l_{\rm N,c}$ of the circular orbits by the relation $d\psi_{\rm N}/dr=l_{\rm N,c}^2/r^3$. 
The pseudo-Newtonian gravitational potential $\psi$ is defined by the Newtonian relation with the transposition $l_{\rm N,c}\rightarrow L_{\rm c}/E_{\rm c} \equiv l_{\rm c}$ \footnote{The quantity $l_{\rm c}=L_{\rm c}/E_{\rm c}$ plays its role only when the pseudo-Newtonian (e.g, P-W) gravitational potential is defined. Later, standard Newtonian quantities in Newtonian theory are used along with the pseudo-Newtonian gravitational potential.}, i.e., by the relation
\begin{eqnarray}
\label{11}
\psi=\int\frac{L_{\rm c}^2}{E_{\rm c}^2\,r^3}dr.
\end{eqnarray}
Note that the described method of pseudo-Newtonian determination works quite well in spherically symmetric (non-rotating) spacetimes, or in the equatorial plane of axially symmetric (rotating, e.g. Kerr or KdS) spacetimes. However, it is much more complicated task to find a pseudo-Newtonian potential for the regions outside the equatorial plane of the rotating spacetimes, because of a non-trivial influence of the dragging of inertial frames. There is a need to upgrade this method\cite{Gho-Muk:2007} or use completely different way of the gravitational potential definition\cite{Sem-Kar:1999}.  

\begin{figure}[tpb]
\centerline{\psfig{file=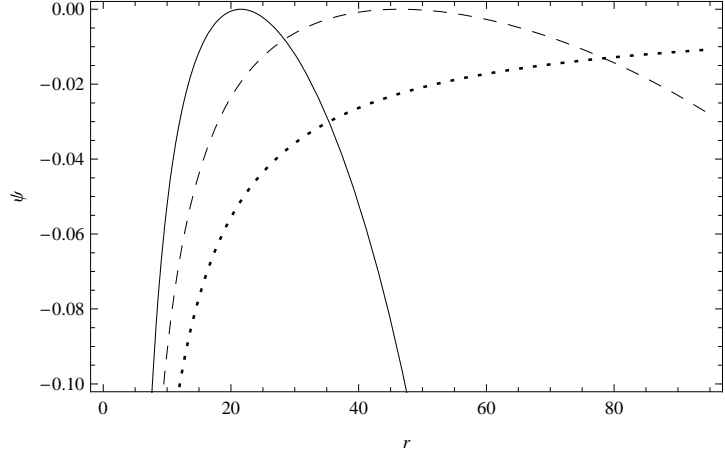,width=0.75\hsize}}
\vspace*{8pt}
\caption{\label{Fig:2}Pseudo-Newtonian gravitational potential $\psi(r;y)$ for three values of the cosmological parameter $y=0$ (dotted), $y=10^{-5}$ (dashed) and $y=10^{-4}$ (solid). The values of $y$ are chosen high enough to exhibit in most illustrative way the structure of SdS spacetimes allowing the existence of accretion discs.}
\end{figure}

In general relativity, the circular orbits in the SdS spacetimes correspond to extrema of the effective potential\cite{Stu-Hle:1999} 
\begin{eqnarray}
\label{7}
V_{\rm eff}(r;L,y)=\left[\left(1-\frac{2}{r}-yr^2\right)\left(1+\frac{L^2}{r^2}\right)\right]^{1/2},
\end{eqnarray}
whereas the related constant of motion take the form 
\begin{eqnarray}
\label{8}
L_{\rm c}(r;y)&\equiv&\left[r(1-yr^3)\right]^{1/2}\left(1-\frac{3}{r}\right)^{-1/2},\\
\label{81}
E_{\rm c}(r;y)&\equiv&\left(1-\frac{2}{r}-yr^2\right)\left(1-\frac{3}{r}\right)^{-1/2}.
\end{eqnarray}
Thus, the gravitational potential (\ref{11}) can be written in the form
\begin{eqnarray}
\label{82}
\psi=\frac{r}{2(2-r+r^3y)}+K,
\end{eqnarray}
with an integrating constant $K$. We calibrate the potential (\ref{82}) by using the condition 
\begin{eqnarray}
\label{3}
\psi(r=r_{\rm s})=0,\quad r_{\rm s}=y^{-1/3},
\end{eqnarray}
where $r_{\rm s}$ is the static radius defined above; the gravitational force, defined in the general relativistic framework of the optical reference geometry\cite{Stu-Hle:1999,Hle:2002} disappears at $r=r_{\rm s}$. Existence of equilibrium position of free particles with covariant energy representing the rest energy 
\begin{eqnarray}
E_{\rm s}(y)\equiv E_{\rm c}(r=r_{\rm s};y)=\left(1-3y^{1/3}\right)^{1/2}
\end{eqnarray}
is possible there.  
Therefore, it is natural to consider the static radius as the \textquoteleft starting point\textquoteright, corresponding to infinity in asymptotically flat spacetime. The proposed pseudo-Newtonian gravitational potential (\ref{82}) then takes the form 
\begin{eqnarray}
\label{12}
\psi(r;y)=\frac{r^3y-3ry^{1/3}+2}{2(1-3y^{1/3})(2-r+r^3y)},
\end{eqnarray}
diverges at the radii of horizons and vanishes at the static radius; for $y=0$ the potential tends asymptotically to zero (see Fig.~\ref{Fig:2}).
%*************************************************************************************8
\section{Test-particle motion}
In the case of central gravitational fields, test-particle motion is confined to central planes and we can choose the equatorial plane for an individual particle. Following the Newtonian physics, the radial equation of the Keplerian equatorial motion can be written in the form
\begin{eqnarray}
\label{12a}
\frac{1}{2}\left(\frac{dr}{dt}\right)^2=e-v_{\rm eff}, 
\end{eqnarray}
where $e$ is the total pseudo-Newtonian energy per particle mass (energy hereafter) and $v_{\rm eff}$ is the pseudo-Newtonian effective potential per particle mass (effective potential in the following) defined by the standard relation
\begin{eqnarray}
\label{13}
v_{\rm eff}(r;l,y)=\psi+\frac{l^2}{2r^2}.
\end{eqnarray}
Here, $\psi(r;y)$ is the pseudo-Newtonian gravitational potential (\ref{12}) and $l$ is the pseudo-Newtonian angular momentum per particle mass (angular momentum hereafter) defined in Section \ref{section 2}.
The circular Keplerian orbits (geodesics) correspond to the effective potential extrema\footnote{Keplerian circular motion can be equivalently given also directly from the pseudo-Newtonian gravitational potential (\ref{12}) using relations for orbital and angular velocitites, and for the angular momentum and energy  
\begin{eqnarray*}
\label{15}
v=\left(r\frac{d\psi}{dr}\right)^{1/2},\; \Omega=\left(\frac{1}{r}\frac{d\psi}{dr}\right)^{1/2},\; l_{\rm c}=\left(r^3\frac{d\psi}{dr}\right)^{1/2},\; e_c=\frac{1}{2}v^2+\psi.
\end{eqnarray*}
However, in some sense, the method of effective potential\cite{Mis-Tho-Whe:1973}, combining the gravitational potential and potential of centrifugal forces, is more general, illustrative and convenient for our case.}. Thus, their radii are governed by the function   
\begin{eqnarray}
\label{8a}
l_{\rm c}^2(r;y)\equiv\frac{r^3(1-r^3y)}{(2-r+r^3y)^2},
\end{eqnarray}
and the corresponding energies (effective potential extrema) are governed by the function 
\begin{eqnarray}
\label{14}
e_{\rm c}(r;y)\equiv\frac{1}{2(1-3y^{1/3})}-\frac{r(r-3)}{2(2-r+r^3y)^2}.
\end{eqnarray}
%---------------------------------------
\subsection{Marginally stable and marginally bound orbits}
\begin{figure}[tpb]
\centerline{\psfig{file=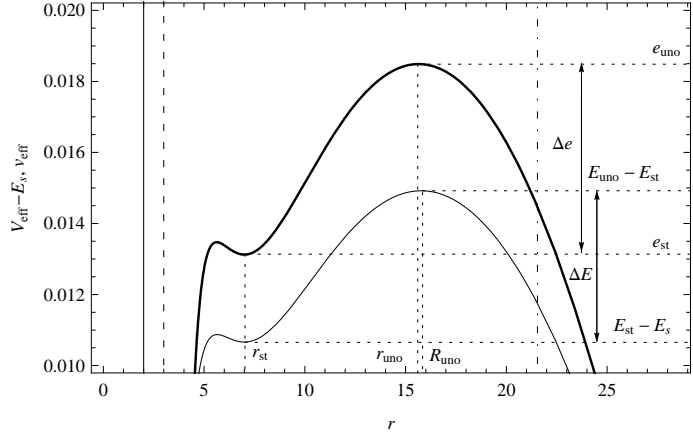,width=0.75\hsize}}
\vspace*{8pt}
\caption{\label{Fig:3}Pseudo-Newtonian $v_{\rm eff}(r;l,y)$ (thick) and relativistic $V_{\rm eff}(r;L,y)-E_{\rm s}(y)$ (thin) effective potentials for test-particle motion with the pseudo-Newtonian $l^2=l^2_{\rm c}(r=7)=13.4)$ and relativistic $L^2=L^2_{\rm c}(r=7)=11.8$ angular momenta in the SdS spacetime with $y=10^{-4}$. Both the potentials determine stable circular orbits at $r=7$. The dotted lines suggest determination of the energy differencies (potential barriers) between the outer unstable orbits, determined in the pseudo-Newtonian and relativistic ways, and the fixed central stable orbit at $r=7$.}
\end{figure}

Behaviour of the pseudo-Newtonian effective potential $v_{\rm eff}$ qualitatively follows  behaviour of its general relativistic counterpart $V_{\rm eff}$ (see Fig.~\ref{Fig:3}).
We demonstrate this statement by comparing behaviour of the functions $l_{\rm c}^2$ and $e_{\rm c}$, and their general relativistic counterparts $L^2_{\rm c}$ and $E_{\rm c}$, defined by  relations (\ref{8}), (\ref{81}), (\ref{8a}) and (\ref{14}). 

\begin{figure}[tpb]
\centerline{\psfig{file=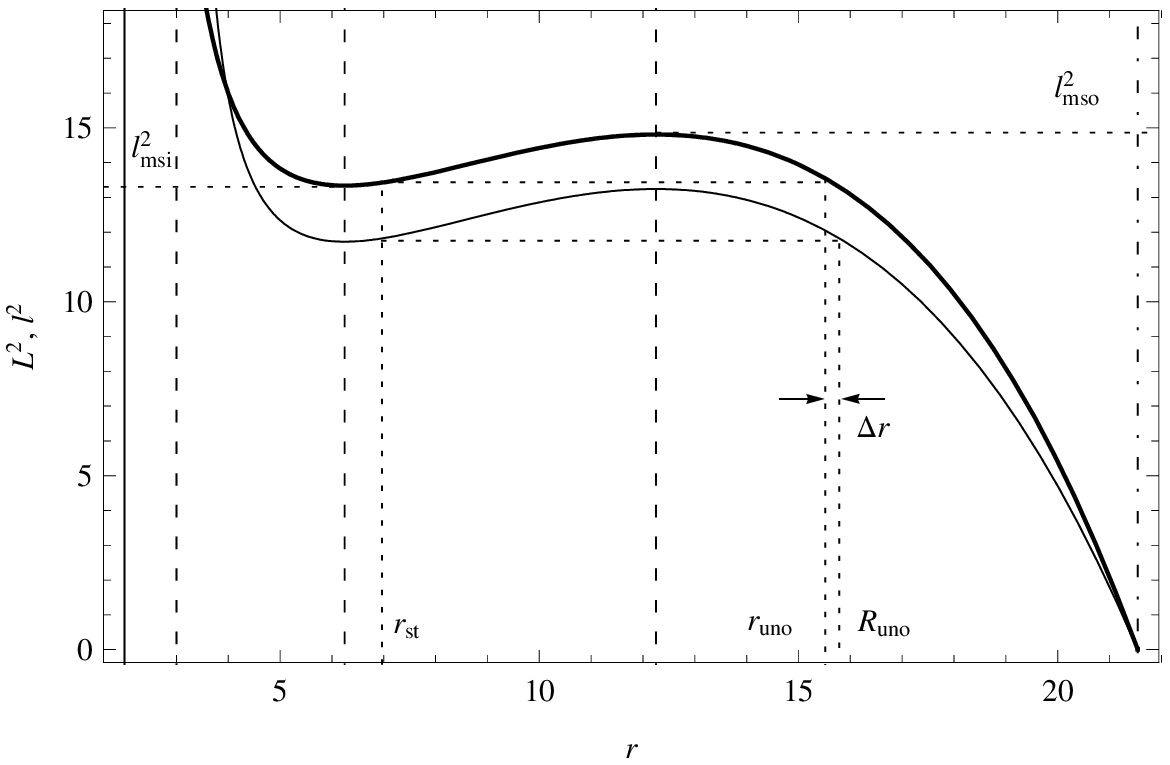,width=0.75\hsize}}
\vspace*{8pt}
\caption{\label{Fig:4}Radial profiles of pseudo-Newtonian $l^2_{\rm c}(r;y)$ (thick) and relativistic $L^2_{\rm c}(r;y)$ angular momenta determining circular orbits (effective potential extrema) in SdS spacetimes with $y=10^{-4}$. We show an example of typical behaviour of the angular momenta, each with two extrema. The solid vertical line denotes the black-hole horizon position; the dashed vertical lines denote photon and marginally stable circular orbits positions; the dashed-dotted vertical line denotes the static radius position. The dotted lines suggest determination of the difference in positions of the outer unstable circular orbits, determined in the pseudo-Newtonian and relativistic ways for fixed position of the central stable orbit at $r=7$.}
\end{figure}

\begin{figure}[tpb]
\centerline{\psfig{file=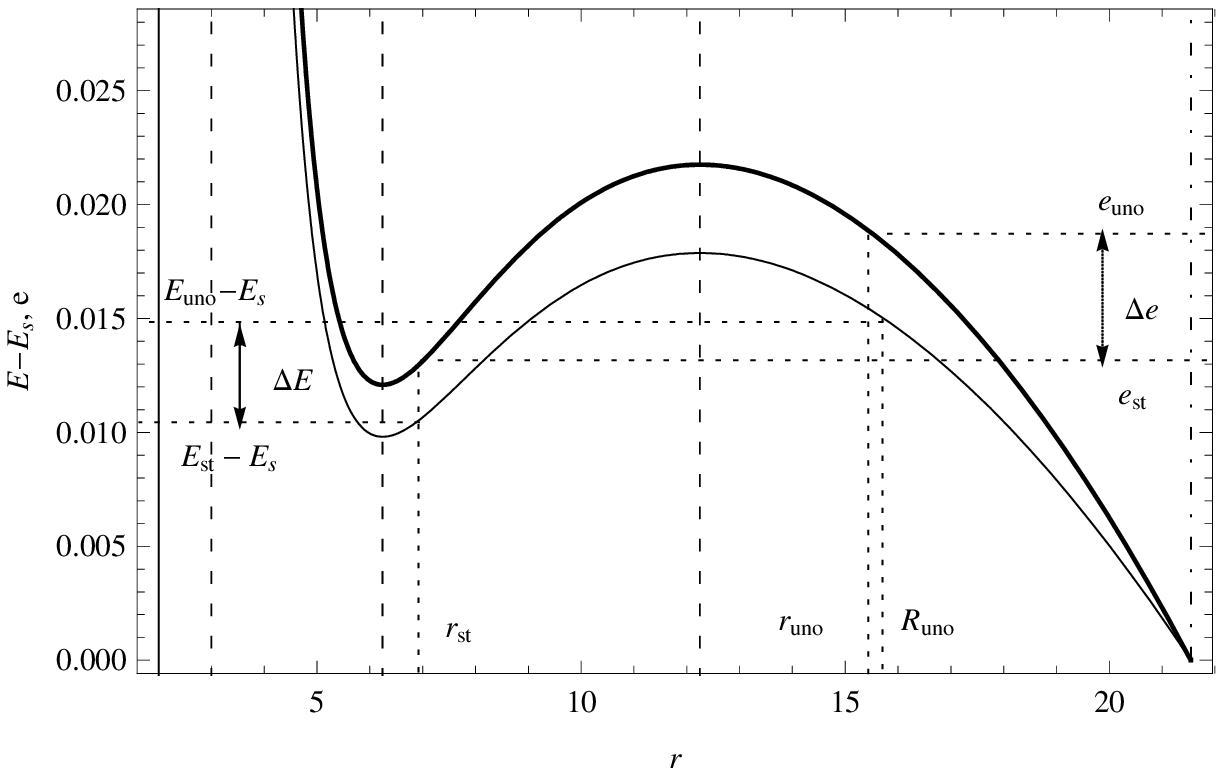,width=0.75\hsize}}
\vspace*{8pt}
\caption{\label{Fig:5}Pseudo-Newtonian $e_{\rm c}(r;y)$ (thick) and relativistic $E_{\rm c}(r;y)-E_{\rm s}(y)$ energies of circular orbits in SdS spacetimes with $y=10^{-4}$. The dotted lines suggest determination of the energy differences (potential barriers) between the outer unstable orbits, determined in the pseudo-Newtonian and relativistic ways, and the central stable orbit at $r=7$ (for the dashed-dotted and dashed vertical lines meaning see Fig.~\ref{Fig:4}).}
\end{figure}

The functions $l^2_{\rm c}$ and $e_{\rm c}$ diverge at the radius of the black-hole horizon $r_{\rm h}$ and vanish at the static radius, while $L^2_{\rm c}$ and $E_{\rm c}$ diverge at the radius of the photon circular orbit, $L^2_{\rm c}$ vanishes at the static radius, and $E_{\rm c}$ does not (see Figs~\ref{Fig:4} and \ref{Fig:5}). 
Thus, the pseudo-Newtonian approach implies that existence of circular orbits is limited by the static radius from above, in accord with the relativistic approach, and by the black-hole horizon from below, in contrast to the relativistic approach, where the lower limit is given by the radius of photon orbit. This is caused by the fact that we do not obtain the photon circular orbit in the pseudo-Newtonian approach\footnote{The same situation occurs for the standard P-W pseudo-Newtonian potential defined for the Schwarzschild spacetimes\cite{Pac-Wii:1980}}.   

As well as in the general relativistic approach, stable circular orbits, determined by the minima of the effective potential $v_{\rm eff}$, satisfy the condition $\partial_r l_{\rm c}^2>0$. The unstable circular orbits, determined by the maxima of $v_{\rm eff}$, satisfy the condition $\partial_r l_{\rm c}^2<0$. Inner and outer marginally stable circular orbits correspond to extrema $l^2_{\rm msi}(y)$ and $l_{\rm mso}^2(y)$ of the function $l_{\rm c}^2$. They are governed by the function  
\begin{eqnarray}
\label{9}
y_{\rm ms}(r)\equiv \frac{r-6}{r^3(4r-15)},
\end{eqnarray}
just as in the general relativistic approach, because loci of extrema of the functions $l^2_{\rm c}$ and $L^2_{\rm c}$ coincide. 
The stable circular orbits are then limited by the condition $4yr^4-15yr^3-r+6\leq0$. 

\begin{figure}[tpb]
\centerline{\psfig{file=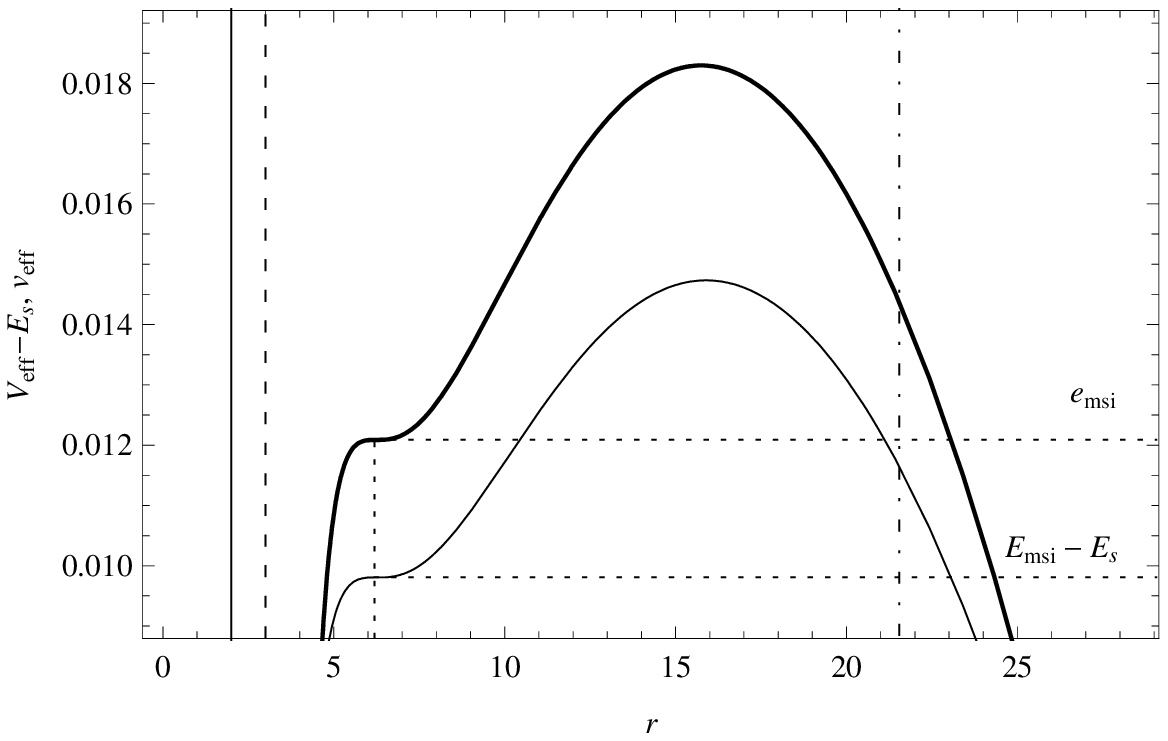,width=0.75\hsize}}
\centerline{\psfig{file=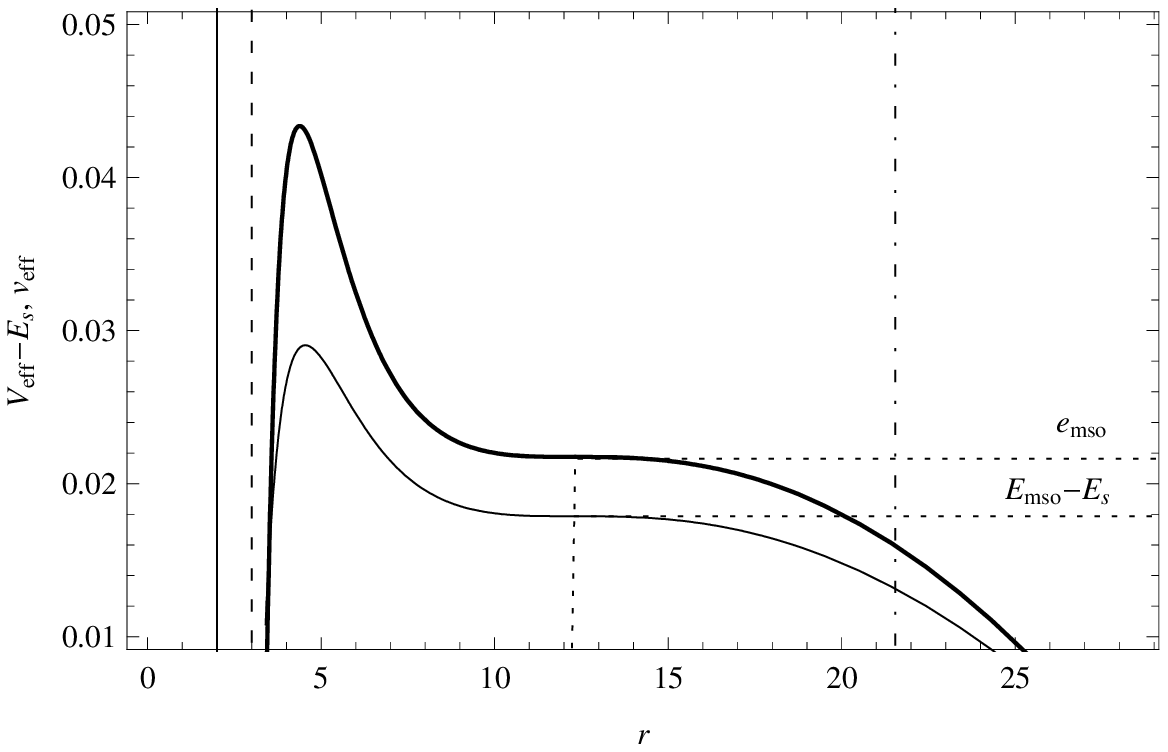,width=0.75\hsize}}
\vspace*{8pt}
\caption{\label{Fig:6}Marginally stable circular orbits. Pseudo-Newtonian $v_{\rm eff}(r;l,y)$ (thick) and relativistic $V_{\rm eff}(r;L,y)-E_{\rm s}(y)$ effective potentials for test-particle motion with the pseudo-Newtonian $l^2=l^2_{\rm c}(r=6.2)=13.3$ (upper figure), $l^2=l^2_{\rm c}(r=12.3)=14.8$ (lower figure), and relativistic $L^2=L^2_{\rm c}(r=6.2)=11.7$ (upper figure), $L^2=L^2_{\rm c}(r=12.3)=13.2$ (lower figure) angular momenta in the SdS spacetime with $y=10^{-4}$. Positions of the marginally stable circular orbits correspond to the positions of the potential inflex points.}
\end{figure}

The function $y_{\rm ms}$ vanishes at the radius $r=6$, corresponding to the marginally stable circular orbit in the Schwarzschild spacetime, diverges at $r=0$ and at $r=15/4$, and has minimum at $r=r_{\rm ph}$, where $y_{\rm ms}(r=r_{\rm ph})=y_{\rm crit}$. The function is irrelevant at the range $0<r<15/4$, where $y_{\rm ms}>y_{\rm crit}$. The physically relevant part of $y_{\rm ms}$ is located at $r\geq6$. Its maximum is located at $r_{\rm ms,e}=15/2$ and the corresponding maximum is $y_{\rm ms}(r=r_{\rm ms,e})\equiv y_{\rm ms,e}=12/15^4$ (see Fig.~\ref{Fig:1}). This value represents the limiting value for the SdS spacetimes admitting stable geodesics, i.e., accretion discs\cite{Stu-Sla-Hle:2000}. Behaviour of the effective potential determining the innermost and outermost stable circular orbits is illustrated in Fig.~\ref{Fig:6}.

The marginally bound circular orbits (inner and outer), corresponding to two unstable circular orbits with the same energy and appropriately chosen angular momentum, are determined by the condition
\begin{eqnarray}
\label{17}
e_{\rm c}(r=r_{\rm mbi};y)=e_{\rm c}(r=r_{\rm mbo};y)\equiv e_{\rm mb}(y).
\end{eqnarray}
\begin{figure}[tpb]
\centerline{\psfig{file=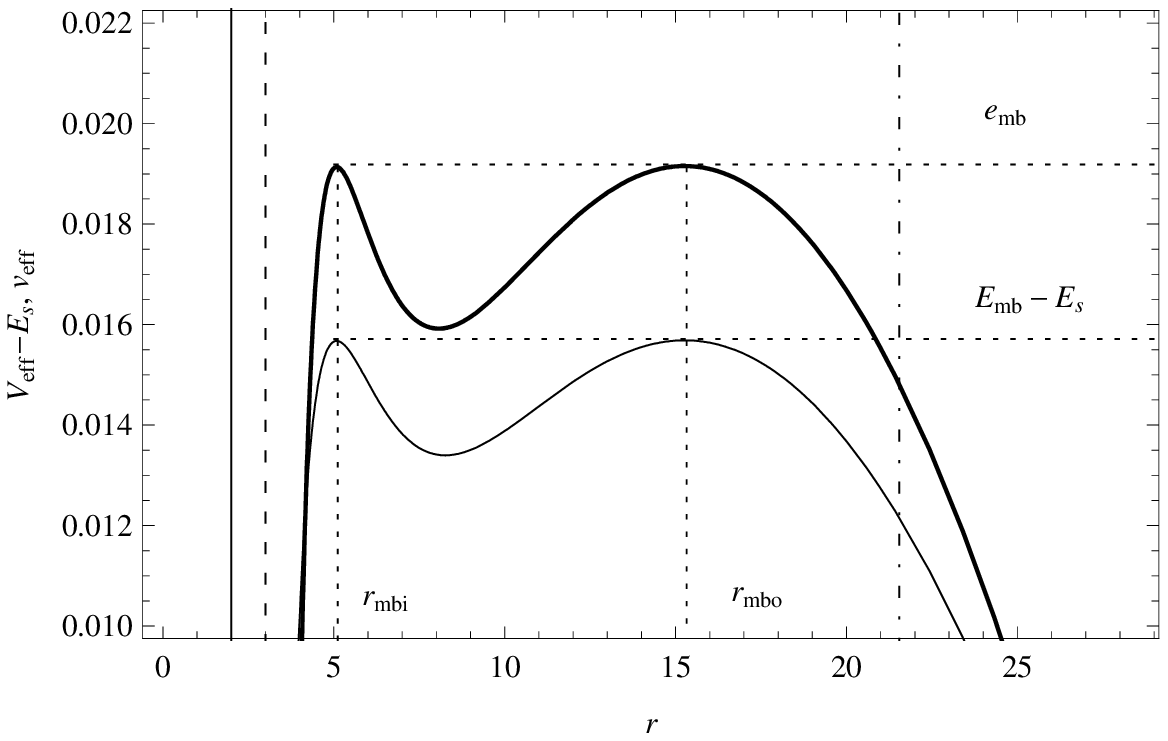,width=0.75\hsize}}
\vspace*{8pt}
\caption{\label{Fig:7}Marginally bound circular orbits. Pseudo-Newtonian $v_{\rm eff}(r;l,y)$ (thick) and relativistic $V_{\rm eff}(r;L,y)-E_{\rm s}(y)$ effective potentials for test-particle motion with the pseudo-Newtonian $l^2=l^2_{\rm c}(r=15.3)=13.7$ and relativistic $L^2=L^2_{\rm c}(r=15.3)=12.2$ angular momenta in the SdS spacetime with $y=10^{-4}$. Positions of the marginally bound circular orbits correspond to positions of effective-potential maxima.}
\end{figure}
The related radii $r_{\rm mbi}$ and $r_{\rm mbo}$ can be determined by using the following numerical procedure.
From equation (\ref{8a}), we express two radii $r_I(l^2_{\rm c};y)$ and $r_{II}(l^2_{\rm c};y)$ corresponding to the inner and outer unstable circular orbits (marginally bound orbits radii candidates). Taking, moreover, the condition (\ref{17}) into account, i.e., solving the equation $e_{\rm c}(r=r_I;y)=e_{\rm c}(r=r_{II};y)$, where $e_{\rm c}(r;y)$ is given by relation (\ref{14}), and considering formula (\ref{17}) once more, we obtain function $y_{\rm mb}(r)$ implicitly determining the radii of the marginally stable orbits $r_{\rm mbi}(y)$ and $r_{\rm mbo}(y)$ (see Fig.~\ref{Fig:7}). General relativistic formulas imply the same results, because 
\begin{eqnarray}
\label{18}
e_{\rm c}(r;y)=\frac{1}{2}\left[E_{\rm s}^{-2}(y)-E_{\rm c}^{-2}(r;y)\right],
\end{eqnarray}
and for fixed values of $r$, corresponding to $r_{\rm mbi}$ and $r_{\rm mbo}$, there is $E_{\rm c}(r=r_{\rm mbi})=E_c(r=r_{\rm mbo})$, and thus $e_c(r=r_{\rm mbi})=e_c(r=r_{\rm mbo})$.
%--------------------------------------------------------------------------
\subsection{Potential barriers}
For values of the cosmological parameter $y$ allowing existence of the stable circular orbits, i.e., $0<y<y_{\rm ms,e}$, the functions $l^2_{\rm c}(r;y)$ and $L^2_{\rm c}(r;y)$ have two extrema governed by the function $y_{\rm ms}(r)$ (see Fig.~\ref{Fig:4}). These extrema correspond to the innermost and outermost stable circular geodesics with $l^2=l^2_{\rm msi}$ and $l^2=l^2_{\rm mso}$, and  $L^2=L^2_{\rm msi}$ and $L^2=L^2_{\rm mso}$, respectively, which determine extension and accretion efficiency of thin, Keplerian discs. The marginally stable perfect fluid configurations with $l^2(r,\theta)=$const and $l^2\in(l^2_{\rm msi},l^2_{\rm mso})$ are determined by intersections of $l^2=$const and $l^2_{\rm c}$ curve that determine the center and edges of the fluid configuration\cite{Stu-Sla-Hle:2000}. Therefore,  we realize a detailed comparison of the potentials $v_{\rm eff}$ and $V_{\rm eff}$, concentrating on the differences in positions of the inner, central and outer circular orbits (determined in the pseudo-Newtonian and relativistic ways), and the differences in the corresponding values of both the potential barriers. 

\begin{figure}[tpb]
\centerline{\psfig{file=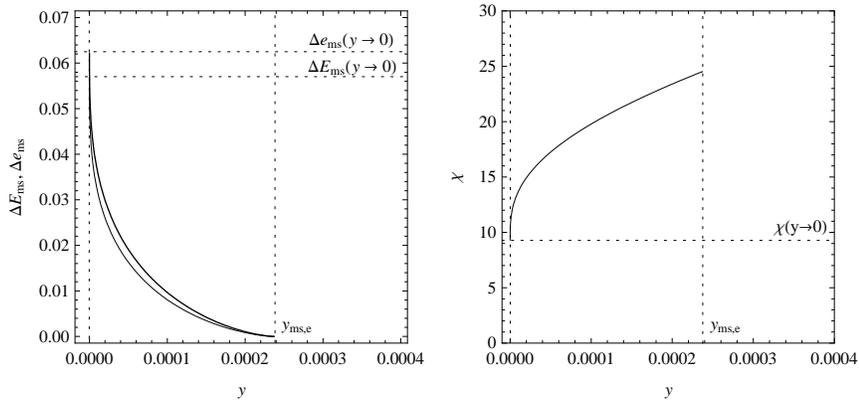,width=0.9\hsize}}
\vspace*{8pt}
\caption{\label{Fig:8}{\bf Left:} Efficiency of the Keplerian discs accretion determined in the pseudo-Newtonian $\Delta e_{\rm ms}(y)$ (thick) and relativistic $\Delta E_{\rm ms}(y)$ ways. {\bf Right:} Characteristic quantity $\chi$ describing the inaccuracy in the determination of the Keplerian accretion efficiency.}
\end{figure}

We start with analyzing the pseudo-Newtonian and relativistic energy differences between the outer and inner marginally stable circular orbits, which determine efficiency of the accretion processes in Keplerian discs  
\begin{eqnarray}
\label{90}
\Delta e_{\rm ms}(y)&=&e_{\rm mso}-e_{\rm msi},\\
\Delta E_{\rm ms}(y)&=&E_{\rm mso}-E_{\rm msi}.
\end{eqnarray}
The energy differences $\Delta e_{\rm ms}$ and $\Delta E_{\rm ms}$ are compared in Fig.~\ref{Fig:8}. Clearly, for astrophysically relevant values\cite{Stu:2002,Stu-Sla-Hle:2000} of $y<10^{-15}$, the pseudo-Newtonian and relativistic efficiencies are very close.
The differences of the pseudo-Newtonian and relativistic efficiency, can be characterized by the quantity  
\begin{eqnarray}
\label{91}
\chi(y)=\frac{\Delta e_{\rm ms} - \Delta E_{\rm ms}}{\Delta E_{\rm ms}}\times100\%,
\end{eqnarray}
the dependence of which on the cosmological parameter is illustrated in Fig.~\ref{Fig:8}. 
In the limit of $y\rightarrow 0$, i.e., in the Schwarzschild spacetimes, there is 
$\Delta e_{\rm ms}\doteq0.063$, $\Delta E_{\rm ms}\doteq0.057$ and $\chi\doteq9.3\%$, and these values hold for the astrophysically relevant values of $y$.

Further, it is of astrophysical relevance to compare dependence of the pseudo-Newtonian and relativistic characteristics of marginally stable discs with uniform distribution of angular momentum $l^2(r,\theta)=$const. Assuming a fixed radius of stable circular orbit corresponding to the disc centre, we can determine and compare the outer edge of the disc and the potential energy well for the matter of the disc. 

\begin{figure}[tpb]
\centerline{\psfig{file=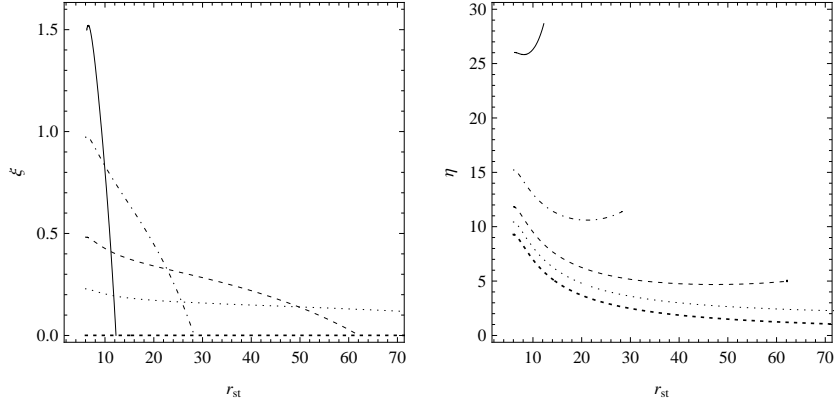,width=0.9\hsize}}
\vspace*{8pt}
\caption{\label{Fig:9}{\bf Left:} Characteristic quantity $\xi(r_{\rm st};y)$ describing the inaccuracies in the pseudo-Newtonian determination of the positions of the outer unstable orbit; {\bf Right:} Characteristic quantity $\eta(r_{\rm st};y)$ describing the inaccuracies in the pseudo-Newtonian determination of the potential barrier between the outer unstable and central stable circular orbits in the SdS spacetimes with $y=10^{-4}$ (solid), $y=10^{-5}$ (dashed-dotted) $y=10^{-6}$ (dashed) $y=10^{-7}$ (dotted) $y=0$ (thick dotted) when position of the stable circular orbit is fixed. The values of $y$ are chosen high enough to exhibit in most illustrative way the influence of $y$ on the calculated quantities.}
\end{figure}

By fixing a value of the radius of stable circular orbit $r=r_{\rm st}$, and calculating the corresponding values of both the pseudo-Newtonian and relativistic angular momenta $l^2_{\rm c}(r=r_{\rm st};y)$ and $L^2_{\rm c}(r=r_{\rm st};y)$ (see Fig.~\ref{Fig:4}), we can determine the radii of the corresponding outer unstable circular geodesics (edge of the disc)  $r_{\rm uno}(r_{\rm st};y)$ (calculated by using $l^2_{\rm c}$) and $R_{\rm uno}(r_{\rm st};y)$ (calculated by using $L^2_{\rm c}$), and their difference
\begin{eqnarray}
\label{92}
\Delta r(r_{\rm st};y)=r_{\rm uno}-R_{\rm uno}.
\end{eqnarray}
The inaccuracies in the pseudo-Newtonian determination of the position of the outer unstable orbit, in dependence on the position of the stable central circular orbit, can be characterized by the quantity  
\begin{eqnarray}
\label{93}
\xi(r_{\rm st};y)=\frac{\Delta r}{R_{\rm uno}-r_{\rm st}}\times100\%.
\end{eqnarray}
We also calculate the pseudo-Newtonian energies at the radius of the stable circular orbit $e_{\rm st}(r_{\rm st};y)\equiv e_c(r=r_{\rm st};y)$ and at the related outer unstable orbit $e_{\rm uno}(r_{\rm st};y)\equiv e_c(r=r_{\rm uno};y)$, and their difference (potential barrier) 
\begin{eqnarray}
\label{94}
\Delta e(r_{\rm st};y)=e_{\rm uno}-e_{\rm st}.
\end{eqnarray}
The same can be done in the relativistic case, i.e., we calculate the energies $E_{\rm st}(r_{\rm st};y)\equiv E_{\rm c}(r=r_{\rm st};y)$ and $E_{\rm uno}(r_{\rm st};y)\equiv E_{\rm c}(r=R_{\rm uno};y)$, and their difference (potential barrier) 
\begin{eqnarray}
\label{95}
\Delta E(r_{\rm st};y)=E_{\rm uno}-E_{\rm st}.
\end{eqnarray}
The inaccuracies in the pseudo-Newtonian determination of such a potential barrier, in dependence on the position of the stable central circular orbit, can be characterized by the quantity 
\begin{eqnarray}
\label{96}
\eta(r_{\rm st};y)=\frac{\Delta e - \Delta E}{\Delta E}\times100\%.
\end{eqnarray}
Dependencies of both the characteristic quantities $\xi$ and $\eta$ are illustrated in Fig.~\ref{Fig:9} for selected values of $y$.  

\begin{figure}[tpb]
\centerline{\psfig{file=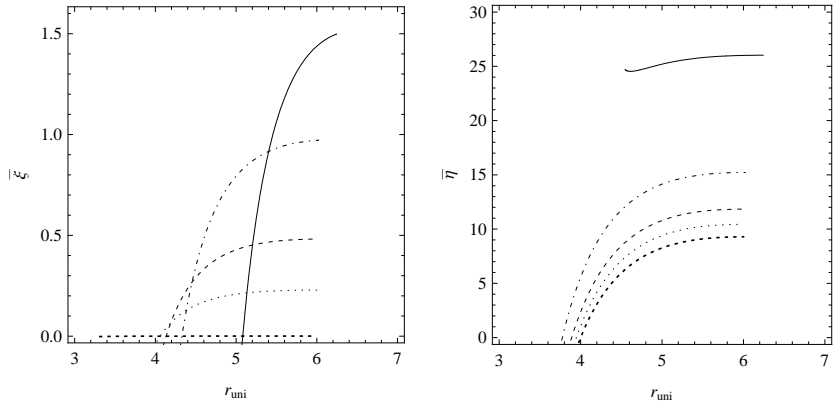,width=0.9\hsize}}
\vspace*{8pt}
\caption{\label{Fig:10}{\bf Left:} Characteristic quantity $\bar{\xi}(r_{\rm uni};y)$ describing the inaccuracies in the pseudo-Newtonian determination of the positions of the outer unstable orbit; {\bf Right:} Characteristic quantity $\bar{\eta}(r_{\rm uni};y)$ describing the inaccuracies in the pseudo-Newtonian determination of the potential barrier between the outer unstable and inner unstable circular orbits in the SdS spacetimes with $y=10^{-4}$ (solid), $y=10^{-5}$ (dashed-dotted) $y=10^{-6}$ (dashed) $y=10^{-7}$ (dotted) $y=0$ (thick dotted), when position of the inner unstable circular orbit is fixed.}
\end{figure}

From the astrophysical point of view, it is better to study the differences between the pseudo-Newtonian and relativistic determination of the radii of the outer unstable circular orbits and potential barriers between the inner and outer unstable orbits, when the position of the inner unstable circular orbit at $r=r_{\rm uni}$ is fixed. In analogy with the previous case, we define the characteristic quantities
\begin{eqnarray}
\label{97}
\bar{\xi}(r_{\rm uni};y)&=&\frac{\Delta \bar{r}}{\bar{R}_{\rm uno}-r_{\rm uni}}\times100\%,\\
\label{98}
\bar{\eta}(r_{\rm uni};y)&=&\frac{\Delta \bar{e}- \Delta \bar{E}}{\Delta \bar{E}}\times100\%,
\end{eqnarray}
where
\begin{eqnarray}
\label{99}
\Delta \bar{r}(r_{\rm uni};y)&=&\bar{r}_{\rm uno}-\bar{R}_{\rm uno},\\
\Delta \bar{e}(r_{\rm uni};y)&=&\bar{e}_{\rm uno}-e_{\rm uni},\\
\Delta \bar{E}(r_{\rm uni};y)&=&\bar{E}_{\rm uno}-E_{\rm uni}.
\end{eqnarray}
The radii $\bar{r}_{\rm uno}(r_{\rm uni};y)$ and $\bar{R}_{\rm uno}(r_{\rm uni};y)$ are now determined by calculating  
$l^2_{\rm c}(r=r_{\rm uni};y)$ and $L^2_{\rm c}(r=r_{\rm uni};y)$, respectively. Corresponding energies are
$\bar{e}_{\rm uno}(r_{\rm uni};y)\equiv e_{\rm c}(r=\bar{r}_{\rm uno};y)$ and $\bar{E}_{\rm uno}(r_{\rm uni};y)\equiv E_{\rm c}(r=\bar{R}_{\rm uno};y)$, and there is $e_{\rm uni}(r_{\rm uni};y)\equiv e_c(r=r_{\rm uni};y)$ and $E_{\rm uni}(r_{\rm uni};y)\equiv E_{\rm c}(r=r_{\rm uni};y)$. Dependencies of both the characteristic quantities $\bar{\xi}$ and $\bar{\eta}$ are given in Fig.~\ref{Fig:10} for selected values of $y$. 

Clearly, the precision in the determination of the radii of the unstable circular orbits and potential barriers, characterized by the quantities  $\xi$ ($\bar{\xi}$) and $\eta$ ($\bar{\eta}$) is high enough, when $y<10^{-6}$. Note that for $y\rightarrow 0$, the positions of unstable circular orbits $r_{\rm uno}(\bar{r}_{\rm uno})\rightarrow \infty$ and $R_{\rm uno} (\bar{R}_{\rm uno})\rightarrow \infty$,  and thus $\xi(\bar{\xi})\rightarrow 0$. In fact, it is irrelevant to study $\xi$ ($\bar{\xi}$) for $y=0$, because there are no outer unstable orbits in the Schwarzschild spacetime. On the other hand, there is  $e_{\rm uno}(\bar{e}_{\rm uno})\rightarrow 0$ and $E_{\rm uno}(\bar{E}_{\rm uno})\rightarrow 1$ for $y\rightarrow 0$ and $r_{\rm uno}(\bar{r}_{\rm uno})\rightarrow \infty$ and $R_{\rm uno}(\bar{R}_{\rm uno})\rightarrow \infty$. Thus, in the case $y=0$, $\eta$ ($\bar{\eta}$) describes the differences in the pseudo-Newtonian and relativistic determinations of the potential barrier between the stable (unstable) orbit and infinity.

%****************************************************************************
\section{Conclusions}
We have shown that the gravitational field of spherically symmetric and static black holes in the universe with a positive cosmological constant, described by the Schwarzschild-de Sitter solution of Einstein's equations, can be alternatively described with a relatively high precision   by using appropriately defined pseudo-Newtonian gravitational potential. 

The presented gravitational potential satisfies important conditions. It admits existence of the static radius, diverges at horizons, and it gives marginally stable and bound orbits at radii exactly equal to those given in the relativistic expressions. The energy difference of these orbits is close to the relativistic relations. 

We have tested the potential correctness by comparing some pseudo-Newtonian results concerning the test-particle geodetical motion with the general relativistic ones. We have been interested in astrophysically relevant situations. Thus, we have chosen range of the cosmological parameter ($0<y\leq 0.000237$), allowing the existence of stable circular geodesics, i.e., existence of accretion discs. 
The differences in the pseudo-Newtonian and relativistic calculations of the accretion processes efficiency have been also studied, whereas the related inaccuracies of the pseudo-Newtonian efficiency determination are less then $12\%$ for $y\leq10^{-6}$, and in astrophysically realistic cases of $y<10^{-15}$, the inaccuracies are very close to those corresponding to the standard P-W potential, describing the Schwarzschild spacetime.    

Moreover, assuming the central circular orbit position fixed, we have compared results of the pseudo-Newtonian and relativistic calculations of the outer unstable circular orbits positions, and the related energy differences (potential barriers) between the central stable (inner unstable) and outer unstable orbits. The comparison suggests that within a certain inaccuracy (less than $0.5\%$ in the case of the radii determination, and less than $12\%$ in the case of the potential barriers determination,  for $y\leq10^{-6}$), the pseudo-Newtonian results are in a good agreement with the relativistic ones. As for the potential barriers, we can state that the differences between the results of both the approaches tend to be smaller for $y$ decreasing and approaching to $y=0$.  

Note that, in principle the existence of three related different kinds of circular orbits, i.e., the inner unstable, central stable and outer unstable orbits, provides us more alternatives for the pseudo-Newtonian gravitational potential testing procedures than presented. We can fix the position of the outer unstable orbit and calculate the differencies in the determination of the central stable orbit and related potential barriers etc. We have not proceeded these comparisons, representing, in a way, alternatives of the presented ones. We prepare a detailed study of the application of the pseudo-Newtonian potential on the perfect fluid dynamics, extending thus further our intention to study accretion processes and the properties of accretion discs in dependence on the cosmic parameter $y$ in the pseudo-Newtonian way. 

\section*{Acknowledgments}
This work was supported by the Czech grants MSM 4781305903 and GA\v{C}R 202/06/0041. One of the authors (Z.S.) would like to express his gratitude to the Czech Committee for Collaboration with CERN for support and Theory Division of CERN for perfect hospitality.

\end{document}